\documentclass[preprint2]{aastex63}

\usepackage[T1]{fontenc}
\usepackage{ae,aecompl}
\usepackage{natbib}

\usepackage{graphicx}	% Including figure files
\usepackage{amsmath}	% Advanced maths commands
\usepackage{amssymb}	% Extra maths symbols

\newcommand\kmsec{\mbox{km~s$^{\rm -1}$}}
\newcommand\species[2]{#1 {\sc #2}}

\shorttitle{The Metamorphosis of the Bailey diagram}
\shortauthors{Bono et al.}

\begin{document}
\title{On the Metamorphosis of the Bailey diagram for RR Lyrae stars}

% generate_latex_authors,['*Bono*','*Brag*','*Cres*','*Fabr*','*Marco*','*Sned*','*Preston*','*Mullen*','*Gillig*','*Fioren*','*Pietrin*','*Altavilla*','*Buonanno*','*Chabo*','*da Silva*','*Ora*','laurainno','*Innocenti*','*Di Carlo*','*Ferraro*','*Grebel*','*Iannico*','*Kovt*','*Kunder*','*Lemasle*','*Marengo*','*Marino*','*Marres*','claramarvaz','*Matsunaga*','*Monelli*','*Neeley*','*Nonino*','*Prada*','*Prudil*','*Stetson*','*venin*','*Togne*','*Valenti*','*Walker*','*Zocca*'],4

% The list of authors, and the short list which is used in the headers.
% If you need two or more lines of authors, add an extra line using \newauthor
\author{G.~Bono }
\affiliation{Dipartimento di Fisica, Universit\`a di Roma Tor Vergata, via della Ricerca Scientifica 1, 00133 Roma, Italy}
\affiliation{INAF-Osservatorio Astronomico di Roma, via Frascati 33, 00078 Monte Porzio Catone, Italy}
\author{V.~F.~Braga }
\affiliation{INAF-Osservatorio Astronomico di Roma, via Frascati 33, 00078 Monte Porzio Catone, Italy}
\affiliation{Space Science Data Center, via del Politecnico snc, 00133 Roma, Italy}
\author{J.~Crestani}
\affiliation{Dipartimento di Fisica, Universit\`a di Roma Tor Vergata, via della Ricerca Scientifica 1, 00133 Roma, Italy}
\affiliation{INAF-Osservatorio Astronomico di Roma, via Frascati 33, 00078 Monte Porzio Catone, Italy}
\affiliation{Departamento de Astronomia, Universidade Federal do Rio Grande do Sul, Av. Bento Gon\c{c}alves 6500, Porto Alegre 91501-970, Brazil}
\author{M.~Fabrizio }
\affiliation{INAF-Osservatorio Astronomico di Roma, via Frascati 33, 00078 Monte Porzio Catone, Italy}
\affiliation{Space Science Data Center, via del Politecnico snc, 00133 Roma, Italy}
\author{C.~Sneden }
\affiliation{Department of Astronomy and McDonald Observatory, The University of Texas, Austin, TX 78712, USA}
\author{M.~Marconi }
\affiliation{INAF-Osservatorio Astronomico di Capodimonte, Salita Moiariello 16, 80131 Napoli, Italy}
\author{G.~W.~Preston}
\affiliation{The Observatories of the Carnegie Institution for Science, 813 Santa Barbara St., Pasadena, CA 91101, USA}
\author{J.~P.~Mullen}
\affiliation{Department of Physics and Astronomy, Iowa State University, Ames, IA 50011, USA}
\author{C.~K.~Gilligan }
\affiliation{Department of Physics and Astronomy, Dartmouth College, Hanover, NBalmerH 03755, USA}
\author{G.~Fiorentino }
\affiliation{INAF-Osservatorio Astronomico di Roma, via Frascati 33, 00078 Monte Porzio Catone, Italy}
\author{A.~Pietrinferni }
\affiliation{INAF-Osservatorio Astronomico d'Abruzzo, Via Mentore Maggini snc, Loc. Collurania, 64100 Teramo, Italy}
\author{G.~Altavilla}
\affiliation{INAF-Osservatorio Astronomico di Roma, via Frascati 33, 00078 Monte Porzio Catone, Italy}
\affiliation{Space Science Data Center, via del Politecnico snc, 00133 Roma, Italy}
\author{R.~Buonanno}
\affiliation{INAF-Osservatorio Astronomico d'Abruzzo, Via Mentore Maggini snc, Loc. Collurania, 64100 Teramo, Italy}
\author{B.~Chaboyer }
\affiliation{Department of Physics and Astronomy, Dartmouth College, Hanover, NH 03755, USA}
\author{R.~da Silva}
\affiliation{INAF-Osservatorio Astronomico di Roma, via Frascati 33, 00078 Monte Porzio Catone, Italy}
\affiliation{Space Science Data Center, via del Politecnico snc, 00133 Roma, Italy}
\author{M.~Dall'Ora }
\affiliation{INAF-Osservatorio Astronomico di Capodimonte, Salita Moiariello 16, 80131 Napoli, Italy}
\author{S.~Degl'Innocenti}
\affiliation{INFN, Sezione di Pisa, Largo Pontecorvo 3, 56127, Pisa, Italy}
\affiliation{Dipartimento di Fisica ``Enrico Fermi'', Universit\`a di Pisa, Largo Pontecorvo 3, 56127, Pisa, Italy}
\author{E.~Di Carlo}
\affiliation{INAF-Osservatorio Astronomico d'Abruzzo, Via Mentore Maggini snc, Loc. Collurania, 64100 Teramo, Italy}
\author{I.~Ferraro}
\affiliation{INAF-Osservatorio Astronomico di Roma, via Frascati 33, 00078 Monte Porzio Catone, Italy}
\author{E.~Grebel}
\affiliation{Astronomisches Rechen-Institut, Zentrum f\"ur Astronomie der Universit\"at Heidelberg, M\"onchhofstr. 12-14, D-69120 Heidelberg, Germany}
\author{G.~Iannicola}
\affiliation{INAF-Osservatorio Astronomico di Roma, via Frascati 33, 00078 Monte Porzio Catone, Italy}
\author{L.~Inno}
\affiliation{Universit\`a  Parthenope di Napoli, Science and Technology Department, CDN IC4, 80143 Naples, Italy}
\affiliation{INAF-Osservatorio Astronomico di Capodimonte, Salita Moiariello 16, 80131 Napoli, Italy}
\author{V.~Kovtyukh}
\affiliation{Astronomical Observatory, Odessa National University, Shevchenko Park, UA-65014 Odessa, Ukraine}
\affiliation{Newton Institute of Chile, Odessa branch, Shevchenko Park, UA-65014 Odessa, Ukraine}
\author{A.~Kunder}
\affiliation{Saint Martin's University, 5000 Abbey Way SE, Lacey, WA 98503, USA}
\author{B.~Lemasle}
\affiliation{Astronomisches Rechen-Institut, Zentrum f\"ur Astronomie der Universit\"at Heidelberg, M\"onchhofstr. 12-14, D-69120 Heidelberg, Germany}
\author{M.~Marengo }
\affiliation{Department of Physics and Astronomy, Iowa State University, Ames, IA 50011, USA}
\author{S.~Marinoni }
\affiliation{INAF-Osservatorio Astronomico di Roma, via Frascati 33, 00078 Monte Porzio Catone, Italy}
\affiliation{Space Science Data Center, via del Politecnico snc, 00133 Roma, Italy}
\author{P.~M.~Marrese }
\affiliation{INAF-Osservatorio Astronomico di Roma, via Frascati 33, 00078 Monte Porzio Catone, Italy}
\affiliation{Space Science Data Center, via del Politecnico snc, 00133 Roma, Italy}
\author{C.~E.~Mart{\'i}nez-V{\'a}zquez}
\affiliation{Cerro Tololo Inter-American Observatory, NSF's National Optical-Infrared Astronomy Research Laboratory, Casilla 603, La Serena, Chile}
\author{N.~Matsunaga }
\affiliation{Department of Astronomy, The University of Tokyo, 7-3-1 Hongo, Bunkyo-ku, Tokyo 113-0033, Japan}
\author{M.~Monelli }
\affiliation{Instituto de Astrof\'isica de Canarias, Calle Via Lactea s/n, E38205 La Laguna, Tenerife, Spain}
\author{J.~Neeley }
\affiliation{Department of Physics, Florida Atlantic University, 777 Glades Rd, Boca Raton, FL 33431 USA}
\author{M.~Nonino }
\affiliation{INAF-Osservatorio Astronomico di Trieste, Via G.~B. Tiepolo 11, 34143 Trieste, Italy}
\author{P.G.~Prada Moroni}
\affiliation{INFN, Sezione di Pisa, Largo Pontecorvo 3, 56127, Pisa, Italy}
\affiliation{Dipartimento di Fisica ``Enrico Fermi'', Universit\`a di Pisa, Largo Pontecorvo 3, 56127, Pisa, Italy}
\author{Z.~Prudil}
\affiliation{Astronomisches Rechen-Institut, Zentrum f\"ur Astronomie der Universit\"at Heidelberg, M\"onchhofstr. 12-14, D-69120 Heidelberg, Germany}
\author{P.~B.~Stetson }
\affiliation{Herzberg Astronomy and Astrophysics, National Research Council, 5071 West Saanich Road, Victoria, British Columbia V9E 2E7, Canada}
\author{F.~Th{\'e}venin}
\affiliation{Universit{\'e} de Nice Sophia-antipolis, CNRS, Observatoire de la C\^{o}te d’Azur, Laboratoire Lagrange, BP 4229, F-06304 Nice, France}
\author{E. Tognelli }
\affiliation{Dipartimento di Fisica ``Enrico Fermi'', Universit\`a di Pisa, Largo Pontecorvo 3, 56127, Pisa, Italy}
\author{E.~Valenti }
\affiliation{European Southern Observatory, Karl-Schwarzschild-Str. 2, 85748 Garching bei Munchen, Germany}
\author{A.~R.~Walker }
\affiliation{Cerro Tololo Inter-American Observatory, NSF's National Optical-Infrared Astronomy Research Laboratory, Casilla 603, La Serena, Chile}
%\author{M.~Zoccali }
%\affiliation{Instituto Milenio de Astrof{\'i}sica, Santiago, Chile}
%\affiliation{Pontificia Universidad Catolica de Chile, Instituto de Astrofisica, Av. Vicu\~na Mackenna 4860, Santiago, Chile}

\begin{abstract}
We collected over 6000 high-resolution spectra of four dozen
field RR Lyrae (RRL) variables pulsating either in the fundamental 
(39 RRab) or in the first overtone (9 RRc) mode. 
We measured radial velocities (RVs) of four strong metallic and 
four Balmer lines along the entire pulsational cycle and derived
RV amplitudes with accuracies better than 1$-$2~\kmsec.
The new amplitudes were combined with literature data for 
23~RRab and 3~RRc stars (total sample 74 RRLs) which allowed us 
to investigate the variation of the Bailey diagram 
(photometric amplitude versus period) when moving from optical to 
mid-infrared bands and to re-cast the Bailey diagram in terms of 
RV amplitudes.  We found that RV amplitudes for RRab are minimally 
affected by nonlinear phenomena (shocks) and multi-periodicity 
(Blazhko effect).  The RV slope ($\log P$--A(V$_r$)) when compared 
with the visual slope ($\log P$--A($V$)) is shallower and the dispersion, 
at fixed period, decreases by a factor of two. 
We constructed homogeneous sets of Horizontal Branch evolutionary models 
and nonlinear, convective  pulsation models of RRLs to constrain 
the impact of evolutionary effects on their pulsation properties. 
Evolution causes, on the Bailey diagram based on RV amplitudes, 
a modest variation in pulsation period and a large dispersion in amplitude. 
The broad dispersion in period of the Bailey diagram is 
mainly caused by variation in RRL intrinsic parameters (stellar mass, 
chemical composition). 
Empirical evidence indicates that RV amplitudes are an optimal diagnostic 
for tracing the mean effective temperature across the RRab instability strip. 
\end{abstract}

\keywords{stars: horizontal-branch --- stars: oscillations --- stars: variables: RR Lyrae --- techniques: radial velocities}

%%%%%%%%%%%%%%%%%%%%%%%%%%%%%%%%%%%%%%%%%%%%%%%%%%

%%%%%%%%%%%%%%%%% BODY OF PAPER %%%%%%%%%%%%%%%%%%

%%%%%%%%%%%%%%%%%%%%%%%%%%%%%%%%%%%%%%%%%%%%%%%%%%%%%%%%%%%%%%%%%%%%%%%%%%%%%%%%%%%%%%%%%%%%%%%%%%%%
%%%%%%%%%%%%%%%%%%%%%%%%%%%%%%%%%%%%%%%%%%%%%%%%%%%%%%%%%%%%%%%%%%%%%%%%%%%%%%%%%%%%%%%%%%%%%%%%%%%%

\section{Introduction}

The RR Lyrae (RRLs) are a very popular group of radial variables, because 
they can be easily identified thanks to the coupling between the pulsation 
periods (0.25$\lesssim$ P $\lesssim$1.0 days) and the shapes of their light 
curves  \citep{dacosta10,monelli17}.

RRLs pulsate in three different flavors: fundamental (RRab), 
first overtone (RRc) and mixed mode (pulsating simultaneously 
in the fundamental and in the first overtone).  
The Bailey diagram \citep[][]{bailey1919}---luminosity amplitude versus 
logarithmic period---is a fundamental plane to identify the RRL pulsation mode. 
In this plane RRc variables cover the short period,
low amplitude domain, and the shape of their amplitude-period relationship
displays either a "bell" or a "hairpin" shape 
\citep{2011rrls.conf....1B,2013AJ....146..119K,2017A&A...599A.125F}.
The RRab variables cover the long-period, large amplitude area and
display a well defined decrease in luminosity amplitude as 
period increases.
 
The Bailey diagram has also been the crossroad of a long-standing astrophysical 
problem, the so-called Oosterhoff dichotomy, identified over 80 years ago \citep{1939Obs....62..104O}. 
RRLs in Globular Clusters (GCs) can be split according to their mean periods in two groups:
the Oosterhoof type I (OoI), with $<P_{ab}>\sim$0.56, 
$<P_c>\sim$0.31 days, and the
Oosterhoof type II (OoII), with $<P_{ab}>\sim$0.66, 
$<P_c>\sim$0.36 days. Subsequent 
investigations identified a similar dichotomic distribution for the RRc, 
albeit with a difference of only $\sim$0.05 days. Spectroscopic studies 
demonstrated that OoI GCs are more metal-rich and have a smaller ratio 
of RRc to total number of RRL than OoII GCs. 
%Ref. 1 
%More recently, 
%\citet{fabrizio2019} found a linear correlation between mean 
%period and metallicity. Thus suggesting that this dichotomy is a 
%natural consequence of the lack of Galactic, metal-intermediate GCs 
%rich in RRLs.
More recently, \citet{fabrizio2019} found a linear correlation between  
mean period and metallicity; thus, suggesting that this dichotomy is a
natural consequence of the lack of Galactic, metal-intermediate GCs
rich in RRLs.

RRL luminosity amplitudes steadily decrease when shifting to longer
wavelengths. However, the amplitude attains an almost constant value at
wavelengths longer than the $K$-band.
This behavior is caused by the fact that luminosity amplitudes in the 
optical bands are mainly driven by temperature variations, while in the 
near-infrared (NIR)/mid-infrared (MIR), they are mainly driven by 
radius variations \citep{madore13,2017ApJ...841...84N}. 

The  Bailey diagram relies on two observables 
that are independent of distance and reddening. The RRL distribution 
in the Bailey diagram depends on their intrinsic parameters (stellar mass, 
luminosity, effective temperature) and chemical composition, but it is 
also affected by degeneracies caused by three different phenomena.    
{\em i)}---Secondary modulations---Empirical evidence suggests 
that a fraction of RRL are affected by a secondary modulation 
\citep[Blazhko effect,][]{blazhko1907}
with a time scale ranging from tens of days to a few years. This fraction 
in large samples (in the Galactic Bulge) ranges from $\sim$6\% for RRc to $\sim$40\% 
for RRab \citep{2018MNRAS.480.1229N,2017MNRAS.466.2602P}. This means that the position 
of an RRL in the Bailey diagram can change by several tenths of a magnitude 
along the Blazhko cycle. The Blazhko effect partially explains why the RRL display, 
at fixed pulsation period and chemical composition, a well defined intrinsic spread 
in luminosity amplitude. 
{\em ii)}---Evolutionary phase---RRLs are low-mass helium 
burning stars and during this evolutionary phase they evolve off the 
Zero-Age-Horizontal-Branch (ZAHB). In their evolution they experience 
changes both in effective temperature and in surface gravity.  These changes 
cause a variation both in pulsation period and in luminosity amplitude. We still 
lack empirical evidence concerning the variation of the position in the Bailey 
diagram directly caused by evolutionary effects.  
{\em iii)}---Nonlinear phenomena---Theory and observations indicate that RRLs 
experience violent shocks approaching maximum compression \citep{preston2019}.  
Shocks are strongly affected by the variation of the efficiency of convective 
motions \citep{bono94b}. This is the reason why RRab amplitudes display a 
steady decrease close to the red edge of the instability strip 
\citep[IS,][]{bono97d,2016AJ....152..170B}.

The main aim of this investigation is to study the physical mechanisms 
affecting the distribution of RRLs in the Bailey diagram.
%%%%%%%%%%%%%%%%%%%%%%%%%%%%%%%%%%%%%%%%%%%%%%%%%%%%%%%%%%%%%%%%%%%%%%%%%%%%%%%%%%%%%%%%%%%%%%%%%%%%
%%%%%%%%%%%%%%%%%%%%%%%%%%%%%%%%%%%%%%%%%%%%%%%%%%%%%%%%%%%%%%%%%%%%%%%%%%%%%%%%%%%%%%%%%%%%%%%%%%%%

\section{Spectroscopic and photometric datasets}

%per contare le rrl con spettri dupont
% awk -F',' '$2>0{print $2,$14}' RRLs_spectracount_200318.csv | grep -c 0
% awk -F',' '$2>0{print $2,$14}' RRLs_spectracount_200318.csv | grep -c1
% awk -F',' '$2>0{print $2,$14}' RRLs_spectracount_200318.csv | grep -c RRd

The largest and most homogeneous sub-sample of high resolution (HR) spectra 
was collected with the Las Campanas Observatory du~Pont echelle spectrograph.
It includes 6,206 HR ($R \equiv \lambda/\delta\lambda$ =35,000)
spectra of 192 RRLs.  Their typical SNR, in the optical regime 
($\lambda$=5,100 \AA), is $\sim$40.

%per contare le rrl con spettri uves
% awk -F',' '($4>0 || $5>0){print $3,$4,$14}' RRLs_spectracount_200318.csv | grep -c 0
% awk -F',' '($4>0 || $5>0){print $3,$4,$14}' RRLs_spectracount_200318.csv | grep -c1
% awk -F',' '($4>0 || $5>0){print $3,$4,$14}' RRLs_spectracount_200318.csv | grep -c RRd

%per contare le rrl con spettri xshooter
% awk -F',' '$9>0 {print $3,$4,$14}' RRLs_spectracount_200318.csv | grep -c 0
% awk -F',' '$9>0 {print $3,$4,$14}' RRLs_spectracount_200318.csv | grep -c1
% awk -F',' '$9>0 {print $3,$4,$14}' RRLs_spectracount_200318.csv | grep -c RRd

%per contare le rrl con spettri harps
% awk -F',' '$6>0 {print $3,$4,$14}' RRLs_spectracount_200318.csv | grep -c 0
% awk -F',' '$6>0 {print $3,$4,$14}' RRLs_spectracount_200318.csv | grep -c1
% awk -F',' '$6>0 {print $3,$4,$14}' RRLs_spectracount_200318.csv | grep -c RRd

%per contare le rrl con spettri feros
% awk -F',' '$3>0 {print $3,$4,$14}' RRLs_spectracount_200318.csv | grep -c 0
% awk -F',' '$3>0 {print $3,$4,$14}' RRLs_spectracount_200318.csv | grep -c1
% awk -F',' '$3>0 {print $3,$4,$14}' RRLs_spectracount_200318.csv | grep -c RRd

These spectra were augmented with 
272 HR ($R$=34,540--107,200, SNR$\sim$20; 68~RRLs) UVES@VLT spectra; 
41 medium--resolution ($R$=4,300-18,000, SNR$\sim$44; 18~RRLs) X-Shooter@VLT spectra; 
166 HR ($R$=115,000, SNR$\sim$10; 5~RRLs) HARPS@3.6m spectra and 
55 HR ($R$=48,000, SNR$\sim$4; 3~RRLs) FEROS@2.2m (ESO, Chile) spectra. 
%

%per calcolare gli snr medi dei vari strumenti
% awk -F',' '$103=="XSHOOTER"{print $74}' allheaders_200402d.csv | grep -v '\-9999' | awk '{x+=$1; next} END{print x/NR}'
% awk -F',' '$103=="HARPS"{print $74}' allheaders_200402d.csv | grep -v '\-9999' | awk '{x+=$1; next} END{print x/NR}'
% awk -F',' '$103=="FEROS"{print $74}' allheaders_200402d.csv | grep -v '\-9999' | awk '{x+=$1; next} END{print x/NR}'
% awk -F',' '($103=="UVES" || $103=="UVES.p95B"){print $74}' allheaders_200402d.csv | grep -v '\-9999' | awk '{x+=$1; next} END{print x/NR}'

%per contare le rrl con spettri salt
% awk -F',' '$10>0 {print $3,$4,$14}' RRLs_spectracount_200318.csv | grep -c 0
% awk -F',' '$10>0 {print $3,$4,$14}' RRLs_spectracount_200318.csv | grep -c1
% awk -F',' '$10>0 {print $3,$4,$14}' RRLs_spectracount_200318.csv | grep -c RRd

Additionally we included HR spectra  from 
SES@STELLA (15, $R$=55,000 , SNR$\sim$35; 1~RRL),  
HRS@SALT (82, $R\sim$40,000, SNR$\sim$50; 70~RRLs), and 
HARPS-N@TNG (10,$R$=115,000, SNR$\sim$40; 4~RRLs).

% numero di variabili per i quali abbiamo spettri (quindi escluse quelle per cui
% abbiamo solo BW)
% awk -F',' '$8!=$12{print $0}' RRLs_spectracount_200318.csv | grep RRab | wc -l
% awk -F',' '$8!=$12{print $0}' RRLs_spectracount_200318.csv | grep RRc | wc -l
% awk -F',' '$8!=$12{print $0}' RRLs_spectracount_200318.csv | grep RRd | wc -l

% numero di calibranti per i quali abbiamo spettri (quindi escluse quelle per cui
% abbiamo solo BW)
% awk -F',' '$8!=$12 && $15==1{print $0}' RRLs_spectracount_200318.csv | grep RRab | wc -l
% awk -F',' '$8!=$12 && $15==1{print $0}' RRLs_spectracount_200318.csv | grep RRc | wc -l
% awk -F',' '$8!=$12 && $15==1{print $0}' RRLs_spectracount_200318.csv | grep RRd | wc -l

% numero di variabili per i quali abbiamo BW
% awk -F',' '$8>0{print $0}' RRLs_spectracount_200318.csv | grep RRab | wc -l
% awk -F',' '$8>0{print $0}' RRLs_spectracount_200318.csv | grep RRc | wc -l
% awk -F',' '$8>0{print $0}' RRLs_spectracount_200318.csv | grep RRd | wc -l

% numero di calibranti per i quali abbiamo solo BW
% awk -F',' '$8==$12 && $15==1{print $0}' RRLs_spectracount_200318.csv | grep RRab | wc -l
% awk -F',' '$8==$12 && $15==1{print $0}' RRLs_spectracount_200318.csv | grep RRc | wc -l
% awk -F',' '$8==$12 && $15==1{print $0}' RRLs_spectracount_200318.csv | grep RRd | wc -l

%
In total we collected 6,572 spectra for 316 RRLs. Among them, 48 RRLs have
a good/optimal coverage of the pulsation cycle (6,226 spectra). We label these 
variables as RV calibrators. 
These data were complemented with RV curves for 26 
field and cluster RRLs available in the 
literature from Baade-Wesselink (BW) analyses
\citep[][]{storm1994b,2003MNRAS.344.1097B}.
We ended up with a sample of 74 (62 RRab, 12 RRc) RV calibrators 
(see Table~1).

The periods from literature are usually precise enough to 
provide good phasing of the RV curves.  However, we derived the 
pulsation period from RV measurements for 2 RRab and 1 RRc. 
We adopted Gaia $G$-band amplitudes \citep{gaia_dr2,marrese2019} transformed 
into the $V$ band by using the equations provided by \citet{2018A&A...616A...4E}. 
We also used $V$-band photometry from ASAS 
\citep{2002AcA....52..397P}, ASAS-SN \citep{2014ApJ...788...48S,2018MNRAS.477.3145J} 
and from Catalina \citep{2015MNRAS.446.2251T,2013ApJ...763...32D,2013ApJ...765..154D}. 

The optical photometry was combined with MIR $W1$-band photometry collected 
with NEOWISE \citep{mainzer2011}.  The MIR light curves have very good 
phase coverage ($\sim$120 phase points), and good photometric precision 
($\sigma_{W1}\le $0.05 mag). 
Optical and MIR light curves were fitted with analytical functions 
(splines, PLOESS, Fourier parameters) following the same approach by 
\citet{2018AJ....155..137B}. The photometric amplitudes are based 
on the analytical fits.  

%%%%%%%%%%%%%%%%%%%%%%%%%%%%%%%%%%%%%%%%%%%%%%%%%%%%%%%%%%%%%%%%%%%%%%%%%%%%%%%%%%%%%%%%%%%%%%%%%%%%
%%%%%%%%%%%%%%%%%%%%%%%%%%%%%%%%%%%%%%%%%%%%%%%%%%%%%%%%%%%%%%%%%%%%%%%%%%%%%%%%%%%%%%%%%%%%%%%%%%%%
\section{Radial velocity measurements}

To measure RVs we chose three \species{Fe}{i} 
lines, from the multiplet~43, at $\lambda$= 4045.81, 4063.59, 4071.74 \AA\ 
and the \species{Sr}{ii} line at $\lambda$= 4077.71 \AA.
These metallic lines were strong enough to be easily 
identified in most spectra even at very low SNR or very high effective temperatures, 
enabling us to trace the whole pulsation cycle using individual spectra, i.e. 
without co-adding spectra. We also used four 
Balmer lines (H$_\alpha$, H$_\beta$, 
H$_\gamma$, H$_\delta$, $\lambda$= 6562.80, 4861.36, 4340.46, 4101.74 \AA).    
The RVs were measured following the same approach employed by 
\citet{2011PASP..123..384F}. 

%___________________________________________________________________________________________
\begin{figure*}
\includegraphics[width=\textwidth]{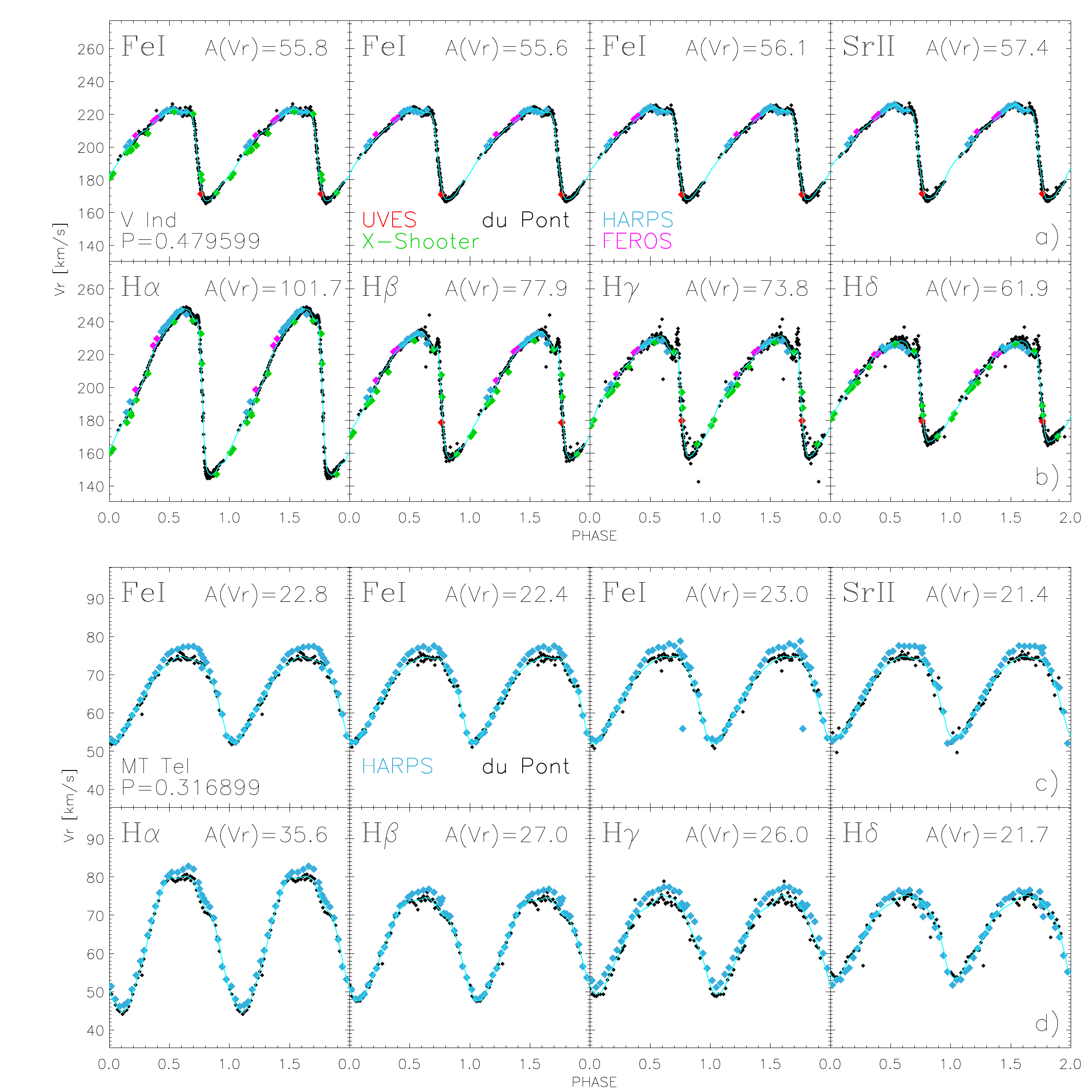}
\caption{Panels a) and b): Radial velocity curves versus phase for the RRab 
V~Ind based on metallic lines \species{Fe}{i} 
($\lambda$: 4045.81, 4063.59, 4071.74 \AA) and \species{Sr}{ii} 
($\lambda$: 4077.71 \AA), 
and on Balmer ($\lambda$: 6562.80, 4861.36, 4340.46, 4101.74 \AA) lines.
Measurements based on different spectrographs are marked with different colors. 
The cyan lines show the analytical fits.
The RV amplitude (A(V$_r$)~\kmsec) and the pulsation period (days) are also 
labeled.
Panels c) and d): Same as the top, but for the RRc MT~Tel.
}\label{fig:Vrad_H_Fe}
\end{figure*}
%___________________________________________________________________________________________

%___________________________________________________________________________________________
\begin{figure}
\includegraphics[width=8.5cm]{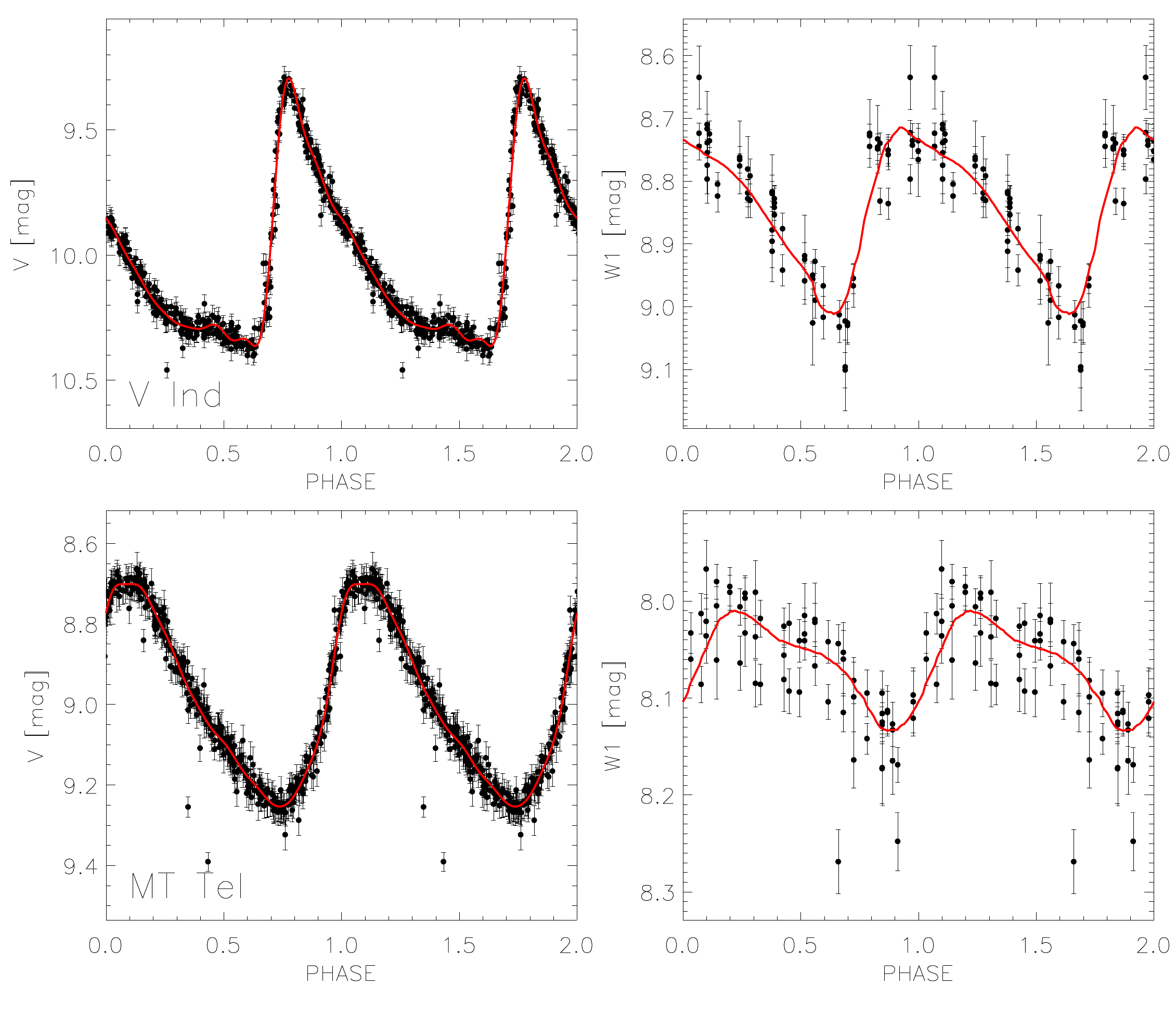}
\caption{Top: Optical ($V$, left) and MIR ($W1$, right) light curves for the RRab V~Ind. 
The optical data were provided by ASAS-SN 
\citep{2014ApJ...788...48S,2018MNRAS.477.3145J}, while the MIR ones 
by NEOWISE \citep{mainzer2011}. The thin red lines show the analytical fits.
Bottom: Same as the top, but for the RRc MT~Tel.}
\label{fig:figura_lcvs}
\end{figure}
%___________________________________________________________________________________________

Figure~\ref{fig:Vrad_H_Fe} shows RV curves for the RRab, V~Ind, and the RRc, MT~Tel.
The RV measurements plotted in this figure display the expected 
difference between metallic and Balmer lines \citep{preston1959}; the RV amplitudes 
for the latter ones are systematically larger (see also Figure~9 of 
\cite{chadid17} for montages of RV and photometric phase plots). 
The difference is caused by the physical conditions at which the different lines 
form:  smaller the optical depth the larger, the RV amplitude.
The difference among H$_\alpha$ and metallic lines ranges from 50\% for MT~Tel 
to almost a factor of two for V~Ind. In spite of the difference in amplitude, 
the secondary features (bumps, dips) appear to be largely co-phased.

We analytically fit the RV curves by using the PLOESS algorithm 
\citep{2018AJ....155..137B}. 
The cyan lines in Figure~\ref{fig:Vrad_H_Fe} show these curves.
Note that RVs for V~Ind based on H$_\gamma$ show 
a well defined ``spike'' around phases 0.70$-$0.76, near its point of minimum
radius, as implied by the optical and MIR light curves of this star in the
top panels of Figure~\ref{fig:figura_lcvs}.

This secondary spike feature was predicted by pulsation models 
(see Figs. 3, 16 in \citealt{bono94b} and Fig.~12 in \citealt{marconi15}) 
suggesting that it is associated with the formation and propagation of shocks. 
Note that the duration of the spike is $\sim$25 minutes and we succeeded in 
tracing it only because the du~Pont spectra were collected with exposure 
times of 400$\pm$200 seconds. Moreover, exposure times for V~Ind are 
$\approx$30\% shorter than the mean, since 
it has a brighter mean visual magnitude (V$=9.92\pm0.05$~mag).
A similar feature is also present among other Balmer (mainly H$_\beta$, 
H$_\delta$) and metallic (mainly Sr) lines, but it is less pronounced. 

Note that the absolute minimum in RV for V~Ind takes place, 
due to the phase lag \citep{castor1971} at the same 
phases of the absolute maximum in the light curves. 
However, the absolute minimum in luminosity is not at the same 
phases of the absolute maximum in RV.  Indeed, the latter 
takes place around the phases in which the optical light curves 
show a well defined bump. The difference appears to be caused by the interplay between 
the shock propagation and the increased efficiency of the convective transport in 
approaching the phases of minimum radius.  

The RV curve of MT~Tel is smoother than that of V~Ind over the entire pulsation cycle. 
For this variable we have extensive measurements from both du~Pont and HARPS
spectrographs. 
Inspection of panels c) and d) of Figure~\ref{fig:Vrad_H_Fe} reveals 
that the RV amplitude from the HARPS dataset is 
$\sim$2$-$3~$\kmsec$ larger than that from du~Pont.
MT~Tel is the only variable for which we found such a difference.
We performed a number of tests that verified the reality of this offset.
It cannot be caused by phasing problems, since the combined optical and 
NIR photometry allowed us very accurate phasing of the two datasets.
The analytical fit displayed in Figure~\ref{fig:Vrad_H_Fe} is mainly 
following the du~Pont measurements, because they have a more uniform 
sampling over the pulsation cycle.
We suggest that the difference is intrinsic and possibly caused by the 
presence of a secondary modulation (Blazhko effect), even though the light 
curves do not show yet a clear modulation at fixed phase. 
Note that HARPS spectra were collected in a single night seven years before 
the du~Pont spectra (over five consecutive nights) and the periodogram  
lacks of relevant secondary peaks.   
This finding would suggest that accurate RV  
measurements might have a stronger sensitivity in detecting secondary 
modulations and in constraining the tomography of variable star atmospheres. 
In the next paper we will explore more aspects of these interesting issues.

% awk -F',' '{print $1" & "$2" & "$3" & "$5" & "$4" & "$6" & "$7" & "$17"$\pm$"$18" & "$19"$\pm$"$24" & "$20"$\pm$"$25" & "$21"$\pm$"$26" & "$22"$\pm$"$27" & "$23"$\pm$"$28}' calibrating_rrls_finaltable_200404.csv

%%%%%%%%%%%%%%%%%%%%%%%%%%%%%%%%%%%%%%%%%%%%%%%%%%%%%%%%%%%%%%%%%%%%%%%%%%%%%%%%%%%%%%%%%%%%%%%%%%%%
%%%%%%%%%%%%%%%%%%%%%%%%%%%%%%%%%%%%%%%%%%%%%%%%%%%%%%%%%%%%%%%%%%%%%%%%%%%%%%%%%%%%%%%%%%%%%%%%%%%%
\section{Results}

%_________________________________________________________________________
\begin{figure}
\includegraphics[width=\columnwidth]{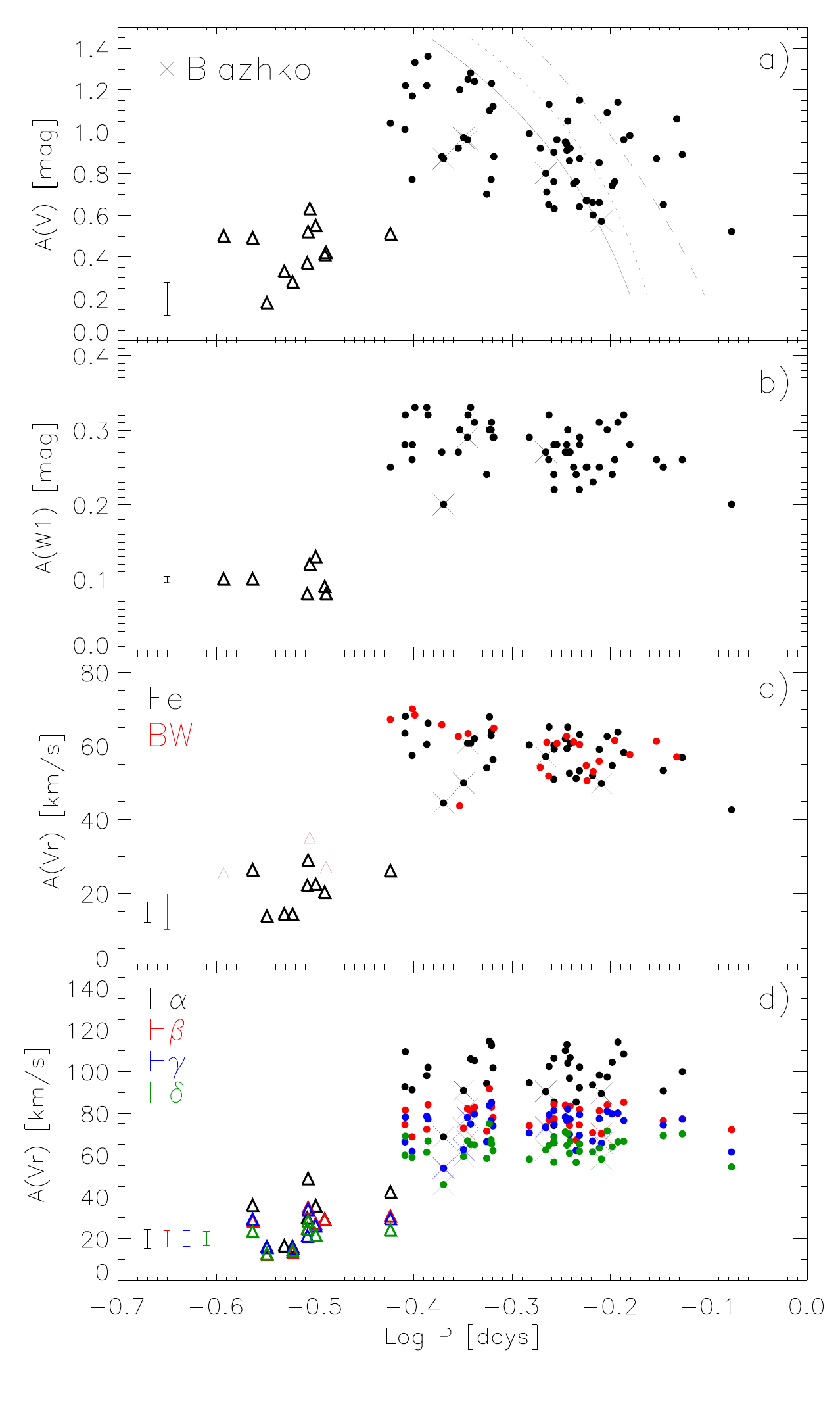}
\caption{Bailey diagram, amplitude versus logarithmic period. 
Panels a) and b) display the luminosity amplitude in the visual 
and in the $W1$ band. 
%Ref. 2 
%The black lines from left to right display 
%observed relations for OoI, Oo-intermediate (dotted) and OoII RRLs 
%provided by \citet{fabrizio2019}.  
The black lines display observed relations for OoI (solid), 
Oo-intermediate (dotted) and OoII (dashed) RRLs provided by 
\citet{fabrizio2019}. 
Panels c) and d) show RV amplitudes based either 
on metallic or on Balmer lines. The red symbols plotted in panel c) display 
BW RRLs. Hydrogen RV amplitudes 
are plotted with different colors (panel d). 
%Ref. 3 green error bar 
Blazhko RRLs are marked with crosses. The error bars plotted in the 
bottom left corner display uncertainties in amplitudes. 
} \label{fig:bailey_vis_mir_vel}
\end{figure}
%_________________________________________________________________________

We have investigated on a more quantitative basis the variation of the
luminosity amplitudes when moving from optical to MIR, and to RV amplitudes
and when moving from weak to strong lines. In Figure~\ref{fig:bailey_vis_mir_vel} 
we show the Bailey diagram for the RRLs for which we estimated both 
luminosity and velocity amplitudes.
Panels a) and b) 
display quite clearly that the luminosity amplitudes in the $W1$-band 
are roughly a factor of three smaller than in $V$-band. 
This trend is expected because the luminosity variation in the 
former band is mainly driven by temperature variations, while in 
the latter it is mainly driven by radius variations. 
Amplitude dispersion at fixed pulsation period also is
smaller in the infrared.
It is of the order of 20\% for the $V$ band ($\sigma_V\sim$0.19 mag), 
while for the $W1$ band it decreases to 10\% ($\sigma_{W1}\sim$0.03 mag).
This outcome applies to both RRab and RRc variables and suggests that the 
dispersion is driven by phenomena affecting the temperature variation 
along the pulsation cycle. Moreover, Blazhko RRLs 
(black crosses) in our sample display a variation in the visual band 
of $\sim$0.4 mag and of $\approx$0.1 mag in the $W1$ band \citep{jurcsik2018}. 

Note that the current sample is quite representative of the RRL 
pulsation properties indeed the analytical relations for 
OoI, Oosterhoff intermediate and OoII RRLs  \citep{fabrizio2019} 
provide a plausible fit of optical amplitudes. 

Figure~\ref{fig:bailey_vis_mir_vel} panels c) and d)
display, for the first time, a Bailey diagram based on RV amplitudes. 
In panel c) together with the current velocity amplitudes we also 
included the velocity amplitudes for field and cluster RRLs (red symbols) 
for which a BW solution is available in the literature.
Interestingly, the relative dispersion in RV amplitudes
is smaller than for visual amplitudes. Indeed, the RV 
amplitudes are purely tracing the radius variation along the pulsation 
cycle. The result is quite clear for RRab for which we find that the 
dispersion is of the order of 10\% ($\sigma_{Vr}\sim$6~\kmsec). 
Moreover, Blazhko RRLs show a variation in velocity 
amplitude significantly smaller ($\sim$20\%) than in visual amplitude 
($\sim$40\%). Unfortunately, we cannot reach firm conclusions concerning 
RRc, since the sample is too small.    

Panel d) of Figure~\ref{fig:bailey_vis_mir_vel}
also shows an interesting trend. The velocity amplitudes 
for RRab variables based on Balmer lines attain an 
almost constant value when moving from the hot (short-period) to the 
cool (long-period) edge of the fundamental IS. The mean
RV amplitudes decrease from H$_\alpha$ to H$_\delta$, 
but their values are constant over the entire period range. 
The dispersion is also modest and ranges from 5 to 11~\kmsec.

%_________________________________________________________________________
\begin{figure*}
\includegraphics[height=18truecm, width=\textwidth]{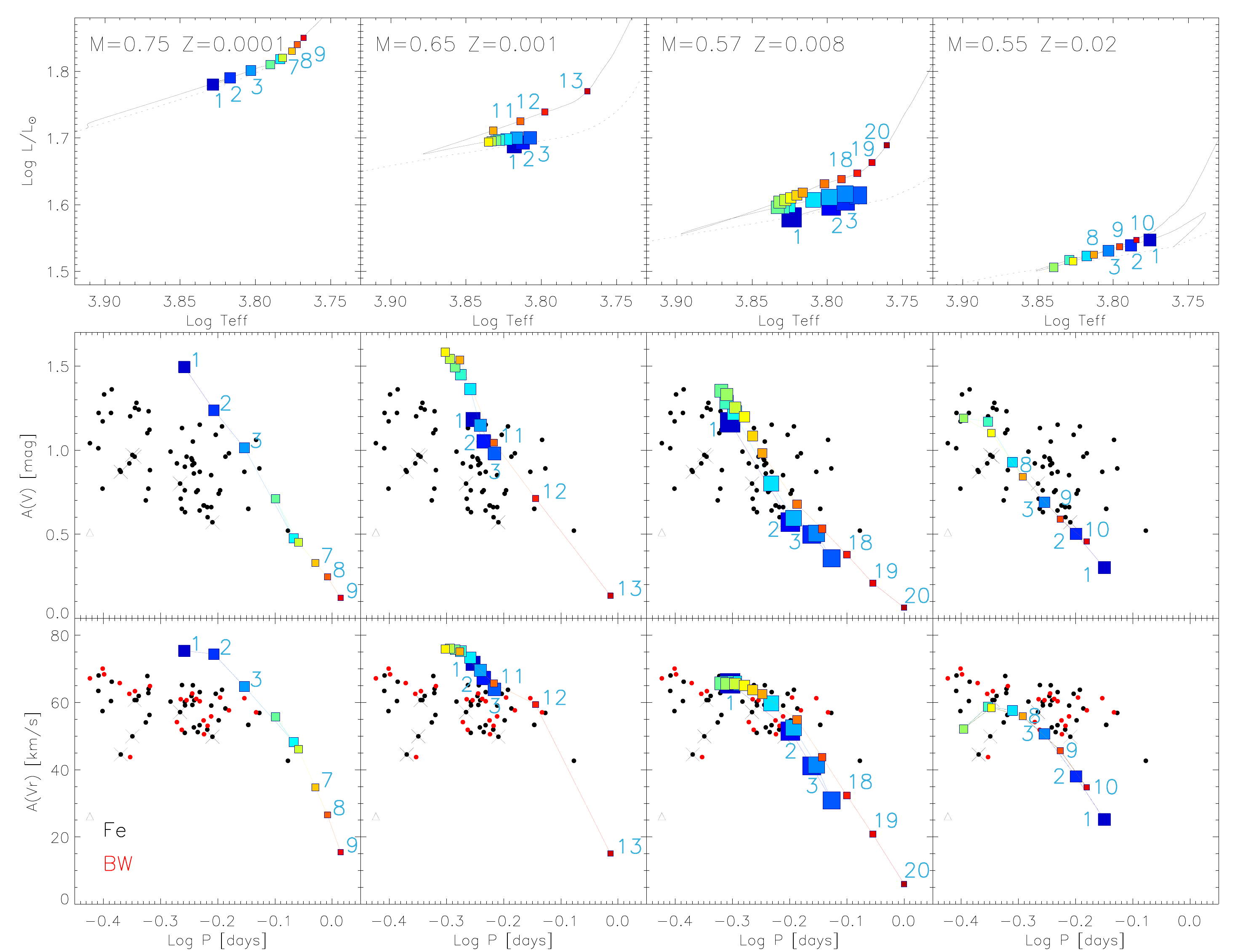}
\caption{Top: Hertzsprung-Russell Diagram for HB models \citep{pietrinferni06b}
constructed by assuming different chemical compositions 
and stellar masses (see labeled values). 
The symbols mark the position of the pulsation models we constructed by using 
evolutionary prescriptions (chemical composition, stellar mass, luminosty, 
effective temperature). 
%Ref. 4a (increased symbol size) and 4b (numbering)  
The light blue numbers mark the first three and the 
last three models in the sequence of pulsation models.    
The symbols are large and bluish close to the ZAHB and become smaller and 
greenish-reddish along the off-ZAHB evolution. 
Middle: Bailey diagram for visual amplitudes. 
The symbols are the same as in panels a) and they are connected with solid 
lines following the same color coding. 
Bottom: Bailey diagram for RV amplitudes based on 
metallic lines. The red symbols display BW RRLs.  
}
\label{fig:RRL_figura_letter_velocity}
\end{figure*}
%_________________________________________________________________________

The RV amplitudes plotted in 
Figure~\ref{fig:bailey_vis_mir_vel} panels c) and d)
clearly show that the dispersion, at fixed period, is significantly 
larger than intrinsic errors; 
% Ref. 5 
thus, suggesting that other physical 
mechanisms can contribute to the observed dispersion. 
To investigate this intrinsic property quantitatively, we 
constructed four HB models covering a broad range in chemical compositions 
(Z=0.0001,0.001,0.008,0.02; Y=0.245,0.245,0.256,0.27). The stellar masses 
(M=0.75,0.65,0.57,0.55 M$_\odot$) were selected to be centrally 
located inside the IS. The reason for this broad range in 
metallicity was driven by the relevant change in the evolutionary path inside 
the IS when moving from metal-poor to metal-rich stellar 
structures. We also constructed, following \citealt{marconi15}, a set 
of non-linear, convective hydrodynamical models of RRL stars along the 
HB evolutionary models to investigate the variation of the pulsation 
properties in their off-ZAHB evolution. 

The top panels of Figure~\ref{fig:RRL_figura_letter_velocity}
show small Hertzsprung-Russell Diagram regions
of the HB evolutionary models (black lines) together with the variable models that are 
pulsationally stable. The color and the size of the symbols change along the off-ZAHB 
evolution.

As expected, the evolution across the IS changes significantly when 
moving from the metal-poor to the metal-rich regime. Indeed, for the most metal-poor 
chemical composition the off-ZAHB evolution is redward, i.e. HB models become 
systematically brighter and cooler as a function of time. The evolutionary path 
becomes more complex at metal-intermediate regime, because the evolutionary 
path undergoes, in sequence, a redward, a blueward and eventually a redward evolution 
approaching its Asymptotic Giant Branch track. This means that we are dealing with triple points, i.e. 
evolutionary phases covering the same range in effective temperature, but slightly 
different luminosities ($\Delta$ Log L/L$_\odot\sim$0.07, Z=0.0001; 
$\sim$0.09, Z=0.001;  $\sim$0.11, Z=0.008; $\sim$0.05, Z=0.02). This means a steady 
decrease in surface gravity, and in turn, a steady increase in pulsation period.  

The change in the physical structure also implies variation in the pulsation 
properties. 
The middle and the bottom panels of 
Figure~\ref{fig:RRL_figura_letter_velocity}
display the comparison between observed and predicted amplitudes. 
This is the first time that evolutionary effects are investigated 
in detail along the off-ZAHB evolution. A glance at the data plotted 
in this figure shows that the models, when moving from metal-poor 
to metal-rich structures, attain, as expected, shorter periods 
and larger amplitudes. 
Interestingly enough, the most metal-poor models display a steady 
decrease in amplitude when moving from hotter to cooler effective 
temperatures.  Note that 
the smaller the symbols the faster is the evolutionary phase. This means that 
the lack of long-period small amplitude RRLs is a mix between the probability 
to find them and the sample size.  

The comparison between theory and observations becomes even more interesting  
in the metal-intermediate and in the metal-rich regime, because pulsation 
models associated with blueward and redward evolutionary phases attain 
very similar values in the 
Bailey diagram. This means that RRLs in their off-ZAHB 
evolution move back and forth in the Bailey diagram covering a modest 
range in period, but a large range in amplitude. This indicates that a 
significant fraction of the dispersion in the Bailey diagram is 
caused by the evolutionary status of RRLs. 
Theory and observations agree quite well over the entire metallicity range. 
Note that predicted amplitudes are affected by uncertainties in 
the mixing length parameter adopted to treat time-dependent convective 
transport in hydrodynamical models of radial variables \citep{bono94b}. 
However, we are using predicted amplitudes in the relative sense, i.e. we are 
interested in the variation of the pulsation amplitudes when moving from the 
blue to the red edge of the IS and not in their absolute value. 

The comparison in the Bailey diagram for RV amplitudes shows similar results. 
Note that to compare theory and observations we adopted a projection factor, 
the ratio between pulsation velocity and RV, p=1.33 \citep{jones92}. 
The comparison in this plane is even more compelling because RV amplitudes 
are marginally affected by nonlinear phenomena and by secondary modulations. 
Interestingly, predicted and observed velocity amplitudes display a quadratic 
variation when moving from the blue to the red edge of the IS, while 
the luminosity amplitudes show a more linear trend.

%%%%%%%%%%%%%%%%%%%%%%%%%%%%%%%%%%%%%%%%%%%%%%%%%%%%%%%%%%%%%%%%%%%%%%%%%%%%%%%%%%%%%%%%%%%%%%%%%%%%
%%%%%%%%%%%%%%%%%%%%%%%%%%%%%%%%%%%%%%%%%%%%%%%%%%%%%%%%%%%%%%%%%%%%%%%%%%%%%%%%%%%%%%%%%%%%%%%%%%%%

\section{Conclusions}

We investigated the properties of the Bailey diagram based on luminosity 
(optical/MIR) and on RV amplitudes. We collected homogeneous RV measurements 
for four dozen of field RRLs.
The current measurements double the sample of RRLs with a good coverage 
of the entire pulsation cycle. We investigated the dependence of luminosity 
amplitudes on nonlinear phenomena, secondary modulations and evolutionary effects. 
The impact of these effects is mitigated when moving from the visual to the MIR bands, 
because the variations along the pulsation cycle are mainly driven by radius than by 
temperature changes. This trend becomes even more compelling in dealing with 
RV amplitudes, indeed, we found that the dispersion, at fixed period, 
is systematically smaller when compared with luminosity amplitudes. Moreover, 
Blazhko RRLs display smaller variations in velocity amplitudes than in 
luminosity amplitudes.    

We constructed four different HB evolutionary models covering a broad range 
in metal contents (Z=0.0001,0.001,0.008,0.02; Y=0.245,0.245,0.256,0.27) and a 
sizable sample of nonlinear, convective hydrodynamical models of RRLs to 
constrain the impact of the off-ZAHB evolution on their pulsation properties. We found that 
evolutionary effects take account for the vertical structure (dispersion in 
amplitude) of the Bailey diagram based on RV amplitudes. 
% Ref. 6 
%Thus suggesting that the dispersion in period is mainly caused by variation 
%in the intrinsic parameters of RRLs (stellar mass, chemical composition). 
Thus, we suggest that the dispersion in period is mainly caused by variation 
in the intrinsic parameters of RRLs (stellar mass, chemical composition).  

The current findings are suggesting that RV amplitudes are not only 
independent of distance and reddening, but also less prone to amplitude 
variations caused by nonlinear phenomena and secondary modulations. 
This means that RV amplitudes are an optimal diagnostic for the variation 
of the mean effective temperature across the IS of RRab.

\section*{Acknowledgements}

We thank an anonumous referee for his/her constructive 
suggestions on an early draft of this paper. 
We acknowledge financial support from 
INAF/ASI 2014-049-R.0 (SSDC), US~NSF under Grants
AST-1714534 (MM, JPM) and AST-1616040 (CS). 
Some of the observations were obtained with SALT. 
This publication uses NEOWISE data 
(JPL/Caltech, University of Arizona, NASA).

%%%%%%%%%%%%%%%%%%%%%%%%%%%%%%%%%%%%%%%%%%%%%%%%%%

\begin{longrotatetable}
\begin{deluxetable*}{lllcccccccccc}
\tabletypesize{\scriptsize}
\tablecaption{Pulsation properties for the current RRLs.}
\tablehead{\colhead{Gaia DR2 ID} & \colhead{Classical ID} & \colhead{type$^a$} & \colhead{Period} & \colhead{A($V$)} & \colhead{A($W1$)} & \colhead{A(Vr)$_{Fe}$} & \colhead{flag$^b$} & \colhead{A(Vr)$_{H\alpha}$} & \colhead{A(Vr)$_{H\beta}$} & \colhead{A(Vr)$_{H\gamma}$} & \colhead{A(Vr)$_{H\delta}$} \\ 
                                 &                        &                    & \colhead{days}   &  \colhead{mag}   &  \colhead{mag}    &  \colhead{\kmsec}      &                    &  \colhead{\kmsec}          &  \colhead{\kmsec}           &  \colhead{\kmsec}     & \colhead{\kmsec}
}  
\startdata
   15489408711727488 &                          X Ari &   0  & 0.65117537 &      0.96 &      0.32 &  58.2$\pm$      2.4 & 0  &     108.2$\pm$      4.4 &      85.1$\pm$      3.3 &      76.4$\pm$      3.0 &      66.5$\pm$      2.9  \\
  234108363683247616 &                         AR Per &   0  & 0.42556048 &      0.88 &      0.27 &  65.7$\pm$      5.1 & 1  &      \ldots             &      \ldots             &      \ldots             &      \ldots              \\
  630421935431871232 &                         RR Leo &   0  & 0.45240000 &      1.25 &      0.32 &  63.3$\pm$      6.8 & 1  &      \ldots             &      \ldots             &      \ldots             &      \ldots              \\
  886793515494085248 &                         RR Gem &   0  & 0.39730000 &      1.17 &      0.28 &  70.0$\pm$     11.3 & 1  &      \ldots             &      \ldots             &      \ldots             &      \ldots              \\
 1009665142487836032 &                         TT Lyn &   0  & 0.59744301 &      0.67 &      0.25 &  50.5$\pm$      3.1 & 1  &      \ldots             &      \ldots             &      \ldots             &      \ldots              \\
 1058066262817534336 &                         SU Dra &   0  & 0.66041178 &      0.98 &      0.28 &  57.6$\pm$      3.5 & 1  &      \ldots             &      \ldots             &      \ldots             &      \ldots              \\
 1191510003353849472 &                         AN Ser &   0  & 0.52207295 &      0.99 &      0.29 &  60.2$\pm$      3.8 & 0  &      94.5$\pm$      6.4 &      73.9$\pm$      5.2 &      70.5$\pm$      4.9 &      57.9$\pm$      3.8  \\
 1286188056265485952 &                         RS Boo &   0  & 0.37736549 &      1.04 &      0.25 &  67.1$\pm$      3.2 & 1  &      \ldots             &      \ldots             &      \ldots             &      \ldots              \\
 1360405567883886720 &            Cl* NGC 6341 SAW V1 &   0  & 0.70279828 &      0.87 &      0.26 &  61.2$\pm$      4.9 & 1  &      \ldots             &      \ldots             &      \ldots             &      \ldots              \\
 1360408905076366208 &           CSS J171712.0+431225 &   0  & 0.63746100 &      0.76 &      0.26 &  61.4$\pm$      3.9 & 1  &      \ldots             &      \ldots             &      \ldots             &      \ldots              \\
 1492230556717187456 &                         TV Boo &   1  & 0.31256000 &      0.63 &      0.12 &  35.0$\pm$      5.6 & 1  &      \ldots             &      \ldots             &      \ldots             &      \ldots              \\
 1683444631037761024 &                         SW Dra &   0  & 0.56967009 &      0.94 &      0.28 &  62.6$\pm$      4.5 & 1  &      \ldots             &      \ldots             &      \ldots             &      \ldots              \\
 1760981190300823808 &                         DX Del &   0  & 0.47261773 &      0.70 &      0.24 &  54.0$\pm$      3.2 & 0  &      94.1$\pm$      6.1 &      71.3$\pm$      4.1 &      66.3$\pm$      4.1 &      58.3$\pm$      3.6  \\
 1793460115244988800 &                         AV Peg &   0  & 0.39038090 &      1.01 &      0.28 &  63.4$\pm$      2.8 & 0  &      92.6$\pm$      4.4 &      74.4$\pm$      3.4 &      66.2$\pm$      3.3 &      59.8$\pm$      3.0  \\
 2381771781829913984 &                         DN Aqr &   0  & 0.63376712 &      0.74 &      0.24 &  54.6$\pm$      2.1 & 0  &     104.3$\pm$      4.0 &      79.6$\pm$      3.4 &      79.6$\pm$      3.5 &      63.8$\pm$      3.0  \\
 2414817603803476864 &                         UU Cet &   0  & 0.60610163 &      0.60 &      0.23 &  53.0$\pm$      3.8 & 1  &      \ldots             &      \ldots             &      \ldots             &      \ldots              \\
 2558296724402139392 &                         RR Cet &   0  & 0.55302505 &      0.90 &      0.28 &  60.1$\pm$      3.1 & 0  &     106.2$\pm$      5.5 &      84.1$\pm$      4.3 &      81.2$\pm$      4.3 &      68.7$\pm$      3.6  \\
 2689556491246048896 &                         SW Aqr &   0  & 0.45930070 &      1.24 &      0.31 &  61.9$\pm$      1.8 & 0  &     105.1$\pm$      3.3 &      82.7$\pm$      2.6 &      79.5$\pm$      2.6 &      64.8$\pm$      2.3  \\
 2720896455287475584 &                         DH Peg &   1  & 0.25551624 &      0.50 &      0.10 &  25.4$\pm$      1.8 & 1  &      \ldots             &      \ldots             &      \ldots             &      \ldots              \\
 2857456207478683776 &                         SW And &   0  & 0.44230000 &      0.92 &      0.27 &  62.5$\pm$      4.6 & 1  &      \ldots             &      \ldots             &      \ldots             &      \ldots              \\
 2981136563930816128 &                         RX Eri &   0  & 0.58725159 &      0.87 &      0.28 &  60.3$\pm$      4.5 & 1  &      \ldots             &      \ldots             &      \ldots             &      \ldots              \\
 3479598373678136832 &                         DT Hya &   0  & 0.56798140 &      0.95 &      0.27 &  61.8$\pm$      3.1 & 0  &     109.9$\pm$      5.7 &      83.7$\pm$      4.3 &      78.1$\pm$      4.2 &      70.9$\pm$      3.7  \\
 3535368455297644928 &            ASAS J110522-2641.0 &   1  & 0.29445920 &      0.33 &  \ldots   &  14.3$\pm$      1.7 & 0  &      16.4$\pm$      2.2 &      \ldots             &      \ldots             &      \ldots              \\
 3546458301374134528 &                          W Crt &   0  & 0.41201190 &      1.36 &      0.32 &  66.1$\pm$      2.5 & 0  &     101.9$\pm$      4.4 &      83.9$\pm$      3.7 &      77.1$\pm$      3.3 &      66.6$\pm$      2.9  \\
 3604450388616968576 &                         AM Vir &   0  & 0.61510000 &      0.66 &      0.25 &  59.0$\pm$      2.5 & 0  &      98.1$\pm$      4.2 &      81.1$\pm$      3.6 &      77.3$\pm$      3.5 &      63.2$\pm$      2.8  \\
 3626569264033312896 &                         AS Vir &   0  & 0.55341000 &      0.63 &      0.22 &  59.1$\pm$      2.0 & 0  &      85.2$\pm$      2.9 &      77.2$\pm$      2.7 &      74.6$\pm$      2.8 &      65.7$\pm$      2.5  \\
 3652665558338018048 &                         ST Vir &   0  & 0.41080754 &      1.22 &      0.33 &  60.3$\pm$      3.4 & 0  &      97.9$\pm$      5.8 &      72.2$\pm$      4.4 &      78.5$\pm$      4.8 &      61.2$\pm$      4.0  \\
 3698725337375614464 &                         UU Vir &   0  & 0.47560267 &      1.10 &      0.30 &  67.8$\pm$      3.1 & 0  &     114.4$\pm$      7.2 &      91.7$\pm$      4.6 &      83.6$\pm$      4.3 &      74.9$\pm$      3.6  \\
 3797319369672686592 &                         SS Leo &   0  & 0.62632619 &      1.09 &      0.30 &  62.5$\pm$      2.0 & 0  &      97.2$\pm$      4.0 &      83.8$\pm$      3.0 &      80.9$\pm$      3.1 &      71.4$\pm$      2.8  \\
 3846786226007324160 &                          T Sex &   1  & 0.32468493 &      0.42 &      0.08 &  27.0$\pm$      2.3 & 1  &      \ldots             &      \ldots             &      \ldots             &      \ldots              \\
 3915998558830693888 &                         ST Leo &   0  & 0.47797595 &      1.23 &      0.31 &  64.0$\pm$      4.3 & 0  &     112.4$\pm$      7.6 &      82.9$\pm$      5.4 &      85.0$\pm$      6.1 &      65.3$\pm$      4.9  \\
 4022618712476736896 &                         TU UMa &   0  & 0.55690000 &      0.96 &      0.28 &  60.6$\pm$      4.3 & 1  &      \ldots             &      \ldots             &      \ldots             &      \ldots              \\
 4055098870077726976 &                      V0494 Sco &   0* & 0.42727100 &      0.87 &      0.20 &  44.5$\pm$      1.7 & 0  &      68.6$\pm$      3.1 &      53.5$\pm$      2.5 &      53.7$\pm$      2.4 &      45.7$\pm$      2.4  \\
 4352084489819078784 &                      V0445 Oph &   0  & 0.39702600 &      0.77 &      0.26 &  57.4$\pm$      4.6 & 0  &      91.1$\pm$      7.1 &      68.6$\pm$      5.1 &      61.6$\pm$      5.1 &      58.8$\pm$      4.8  \\
 4417888542753226112 &                         VY Ser &   0  & 0.71410000 &      0.65 &      0.25 &  53.3$\pm$      4.3 & 0  &      90.6$\pm$      6.3 &      76.3$\pm$     11.2 &      74.2$\pm$      6.5 &      69.2$\pm$      6.0  \\
 4421571803630954752 &           CSS J151841.9+020232 &   0  & 0.54625080 &      0.65 &      0.26 &  51.8$\pm$      3.8 & 1  &      \ldots             &      \ldots             &      \ldots             &      \ldots              \\
 4421571803630958848 &           Cl* NGC 5904 SAW V28 &   0  & 0.54383658 &      0.71 &  \ldots   &  60.9$\pm$      4.7 & 1  &      \ldots             &      \ldots             &      \ldots             &      \ldots              \\
 4433070255716036864 &            ASAS J162158+0244.5 &   1  & 0.32362580 &      0.41 &      0.09 &  20.2$\pm$      2.3 & 0  &      29.2$\pm$      3.2 &      29.0$\pm$      3.8 &      \ldots             &      \ldots              \\
 4454183799545435008 &                         AT Ser &   0  & 0.74655408 &      0.89 &      0.26 &  56.8$\pm$      4.1 & 0  &      99.8$\pm$      7.8 &      77.2$\pm$      5.6 &      77.0$\pm$      5.7 &      70.0$\pm$      5.4  \\
 4467433017735910912 &                         VX Scl &   0  & 0.45518030 &      1.28 &      0.33 &  60.6$\pm$      3.9 & 0  &     105.9$\pm$      7.1 &      81.0$\pm$      5.3 &      74.7$\pm$      5.0 &      64.9$\pm$      4.5  \\
 4596935593202765184 &                         TW Her &   0  & 0.39959577 &      1.33 &      0.33 &  68.3$\pm$      4.7 & 1  &      \ldots             &      \ldots             &      \ldots             &      \ldots              \\
 4689637956899105792 &             Cl* NGC 104 SAW V9 &   0  & 0.73720048 &      1.06 &  \ldots   &  57.0$\pm$      5.3 & 1  &      \ldots             &      \ldots             &      \ldots             &      \ldots              \\
 4709830423483623808 &                          W Tuc &   0  & 0.64224028 &      1.14 &      0.31 &  63.7$\pm$      2.9 & 0  &     114.0$\pm$      5.8 &      79.9$\pm$      4.1 &      80.0$\pm$      4.0 &      66.2$\pm$      3.2  \\
 4860671839583430912 &                         SX For &   0  & 0.60534530 &      0.66 &  \ldots   &  51.9$\pm$      2.9 & 0  &      93.5$\pm$      5.9 &      70.6$\pm$      4.3 &      66.6$\pm$      3.8 &      61.4$\pm$      3.7  \\
 4947090013255935616 &                         CS Eri &   1  & 0.31133020 &      0.52 &  \ldots   &  28.9$\pm$      4.0 & 0  &      48.6$\pm$      3.9 &      34.7$\pm$      3.0 &      33.8$\pm$      2.9 &      28.2$\pm$      2.6  \\
 5022411786734718208 &                         SV Scl &   1  & 0.37735860 &      0.51 &  \ldots   &  26.1$\pm$      2.1 & 0  &      42.2$\pm$      3.2 &      30.6$\pm$      2.6 &      29.4$\pm$      2.6 &      24.0$\pm$      2.6  \\
 5151789464548893184 &                         RV Cet &   0* & 0.61802300 &      0.57 &  \ldots   &  49.8$\pm$      2.3 & 0  &      89.3$\pm$      4.3 &      70.1$\pm$      3.3 &      65.6$\pm$      3.0 &      57.9$\pm$      2.7  \\
 5412243359495900928 &                         CD Vel &   0  & 0.57350788 &      0.86 &      0.27 &  52.5$\pm$      1.8 & 0  &      96.6$\pm$      4.1 &      74.0$\pm$      2.0 &      69.8$\pm$      2.8 &      60.7$\pm$      2.5  \\
 5461994297841116160 &                         WY Ant &   0  & 0.57434364 &      0.92 &      0.27 &  60.6$\pm$      2.9 & 0  &     106.5$\pm$      5.6 &      83.3$\pm$      4.4 &      77.1$\pm$      4.2 &      66.5$\pm$      3.7  \\
 5510293236607430656 &                         HH Pup &   0  & 0.39081190 &      1.22 &      0.32 &  67.9$\pm$      2.6 & 0  &     109.3$\pm$      4.3 &      81.4$\pm$      3.3 &      78.1$\pm$      3.4 &      68.9$\pm$      2.9  \\
 5707380936404638336 &                         BB Pup &   0  & 0.48055043 &      0.88 &      0.29 &  64.8$\pm$      4.8 & 1  &      \ldots             &      \ldots             &      \ldots             &      \ldots              \\
 5768557209320424320 &                         UV Oct &   0* & 0.54257500 &      0.80 &      0.27 &  57.1$\pm$      1.1 & 0  &      90.3$\pm$      2.0 &      73.5$\pm$      1.6 &      73.0$\pm$      1.8 &      62.3$\pm$      1.7  \\
 5769986338215537280 &                         RV Oct &   0  & 0.57117800 &      1.05 &      0.30 &  65.1$\pm$      2.3 & 0  &     103.9$\pm$      3.6 &      82.7$\pm$      2.9 &      81.9$\pm$      3.0 &      70.1$\pm$      2.7  \\
 5773390391856998656 &                         XZ Aps &   0  & 0.58726739 &      1.15 &      0.29 &  63.0$\pm$      1.8 & 0  &     102.0$\pm$      3.2 &      81.7$\pm$      2.5 &      79.3$\pm$      2.7 &      65.1$\pm$      2.5  \\
 5806921716937210496 &                         BS Aps &   0  & 0.58256590 &      0.76 &      0.24 &  51.1$\pm$      1.7 & 0  &      85.2$\pm$      3.2 &      67.1$\pm$      2.5 &      62.0$\pm$      2.3 &      56.4$\pm$      2.3  \\
 5947570591534602240 &                          S Ara &   0* & 0.45184650 &      0.96 &      0.29 &  60.6$\pm$      2.0 & 0  &      \ldots             &      82.1$\pm$      3.0 &      77.9$\pm$      2.8 &      66.7$\pm$      2.5  \\
 6045464303643755008 &            Cl* NGC 6121 SAW V2 &   0  & 0.53568150 &      0.92 &  \ldots   &  54.1$\pm$      6.5 & 1  &      \ldots             &      \ldots             &      \ldots             &      \ldots              \\
 6045485228725626752 &           Cl* NGC 6121 SAW V32 &   0  & 0.57909353 &      0.75 &      0.25 &  61.0$\pm$      6.8 & 1  &      \ldots             &      \ldots             &      \ldots             &      \ldots              \\
 6045491928874778624 &           Cl* NGC 6121 SAW V33 &   0  & 0.61484317 &      0.85 &      0.31 &  55.8$\pm$      8.3 & 1  &      \ldots             &      \ldots             &      \ldots             &      \ldots              \\
 6045502305516138368 &           Cl* NGC 6121 SAW V15 &   0  & 0.44379722 &      1.20 &      0.30 &  43.7$\pm$      6.0 & 1  &      \ldots             &      \ldots             &      \ldots             &      \ldots              \\
 6473212809741024256 &            ASAS J200431-5352.3 &   1  & 0.30022700 &      0.28 &  \ldots   &  14.2$\pm$      1.9 & 0  &      16.1$\pm$      1.9 &      13.2$\pm$      2.2 &      15.4$\pm$      2.2 &      14.1$\pm$      2.3  \\
 6483680332235888896 &                          V Ind &   0  & 0.47959915 &      1.12 &      0.29 &  56.2$\pm$      1.2 & 0  &     101.7$\pm$      2.6 &      77.9$\pm$      2.0 &      73.8$\pm$      2.0 &      61.9$\pm$      1.7  \\
 6526559499016401408 &                         RV Phe &   0  & 0.59641862 &      0.67 &      0.25 &  54.6$\pm$      4.8 & 1  &      \ldots             &      \ldots             &      \ldots             &      \ldots              \\
 6541769554459131648 &                         BO Gru &   1  & 0.28272000 &      0.18 &  \ldots   &  13.7$\pm$      1.5 & 0  &      16.0$\pm$      2.3 &      12.2$\pm$      2.3 &      15.6$\pm$      2.6 &      12.7$\pm$      2.6  \\
 6570585628216929408 &                         TY Gru &   0  & 0.57006515 &      0.91 &      0.27 &  59.2$\pm$      2.1 & 0  &     112.8$\pm$      4.7 &      82.5$\pm$      3.3 &      76.2$\pm$      3.2 &      64.6$\pm$      3.2  \\
 6625215584995450624 &                         AE PsA &   0  & 0.54674000 &      1.13 &      0.32 &  65.1$\pm$      6.5 & 0  &     102.3$\pm$      8.7 &      76.6$\pm$      7.0 &      79.1$\pm$      7.6 &      64.5$\pm$      6.2  \\
 6662886605712648832 &                         MT Tel &   1  & 0.31689945 &      0.55 &      0.13 &  22.4$\pm$      1.4 & 0  &      35.6$\pm$      1.9 &      27.0$\pm$      1.6 &      26.0$\pm$      2.3 &      21.7$\pm$      1.9  \\
 6680420204104678272 &                      V1645 Sgr &   0  & 0.55292000 &      0.76 &      0.24 &  50.9$\pm$      1.9 & 0  &      74.0$\pm$      2.8 &      66.2$\pm$      2.5 &      65.8$\pm$      2.7 &      56.5$\pm$      2.4  \\
 6701821205809488384 &            ASAS J181215-5206.9 &   0  & 0.83753980 &      0.52 &      0.20 &  42.6$\pm$      5.6 & 0  &      \ldots             &      72.0$\pm$      9.5 &      61.3$\pm$      8.2 &      54.2$\pm$      7.4  \\
 6771307454464848768 &                      V0440 Sgr &   0  & 0.47750000 &      0.77 &      0.30 &  62.7$\pm$      5.1 & 0  &     113.2$\pm$      8.9 &      84.0$\pm$      6.5 &      76.5$\pm$      6.4 &      67.1$\pm$      5.2  \\
 6787617919184986496 &                          Z Mic &   0  & 0.58692775 &      0.64 &      0.22 &  53.2$\pm$      1.8 & 0  &      92.1$\pm$      3.7 &      74.3$\pm$      3.0 &      69.3$\pm$      2.8 &      61.7$\pm$      2.6  \\
 6856027093125912064 &            ASAS J203145-2158.7 &   1  & 0.31071060 &      0.37 &      0.08 &  22.0$\pm$      2.0 & 0  &      29.7$\pm$      2.8 &      24.3$\pm$      2.6 &      21.1$\pm$      2.5 &      24.6$\pm$      4.2  \\
 6883653108749373568 &                         RV Cap &   0* & 0.44774990 &      0.97 &  \ldots   &  49.9$\pm$      2.0 & 0  &      90.8$\pm$      3.6 &      72.7$\pm$      3.1 &      62.4$\pm$      2.7 &      59.2$\pm$      2.7  \\
 6884361748289023488 &                         YZ Cap &   1  & 0.27345290 &      0.49 &      0.10 &  26.3$\pm$      2.1 & 0  &      35.8$\pm$      3.0 &      28.3$\pm$      2.8 &      29.1$\pm$      2.9 &      23.2$\pm$      2.6  \\
  \enddata
\tablenotetext{a}{Pulsation mode: 0=RRab; 1=RRc. Candidate Blazhko RRLs are marked with an asterisk.}
\tablenotetext{b}{RV amplitudes: 0=current investigation; 1=BW sample.}
\end{deluxetable*}
\end{longrotatetable}


\bibliographystyle{aa}
% \bibliography{baileyletter_chris} % if your bibtex file is called example.bib

%%%%%%%%%%%%%%%%%%%%%%%%%%%%%%%%%%%%%%%%%%%%%%%%%%

% Don't change these lines
\label{lastpage}
\end{document}